\documentclass[conference]{IEEEtran}
\IEEEoverridecommandlockouts

\usepackage{ulem}
\usepackage{cite}
\usepackage{amsmath,amssymb,amsfonts}
\usepackage{algorithm}
\usepackage{algorithmicx}
\usepackage{algpseudocode}
\usepackage{graphicx}
\usepackage{textcomp}
\usepackage{xcolor}
\usepackage[hyphens]{url}
\usepackage{graphicx}
\usepackage{subcaption}
\usepackage{booktabs}
\usepackage{pifont}
\usepackage[hidelinks]{hyperref}
\usepackage[htt]{hyphenat}
\usepackage{cleveref}
\usepackage{tabularx}
\usepackage{array}
\usepackage{listings}

\lstdefinestyle{mystyle}{
    backgroundcolor=\color{gray!10},
    commentstyle=\color{green!70!black},
    keywordstyle=\color{blue}, 
    numberstyle=\tiny\color{gray},
    stringstyle=\color{purple},
    basicstyle=\ttfamily\footnotesize,
    breakatwhitespace=false,         
    breaklines=true,
    captionpos=b,
    keepspaces=true,                 
    numbers=left,
    numbersep=5pt,
    xleftmargin=1.5em,
    showspaces=false,                
    showstringspaces=false,
    showtabs=false,                  
    tabsize=2,
    language=Python
}

\newcolumntype{R}{>{\raggedleft\arraybackslash}X}
\newcolumntype{C}{>{\centering\arraybackslash}X}

\usepackage{caption}
\DeclareCaptionFont{mybold}{\bfseries}
\DeclareCaptionFont{myninept}{\fontsize{8pt}{8pt}\selectfont}
\IEEEaftertitletext{\vspace{-1.5\baselineskip}}

\usepackage{siunitx}
\usepackage{colortbl}
\usepackage{xspace}

\NewDocumentCommand{\var}{O{s} m O{}}{%
  \ensuremath{#1_{#2}^{#3}}
}

\newcommand{\commentout}[1]{}

\definecolor{light-gray}{gray}{0.80}

\newcommand\aref{Alg.~\ref}
\newcommand\alref{Line~\ref}
\newcommand\appref{Appx.~\ref}
\newcommand\eref{Eq.~\ref}
\newcommand\fref{Fig.~\ref}
\newcommand\tref{Tab.~\ref}
\newcommand\sref{\S~\ref}

\usepackage{amsthm}

\usepackage{amsthm}

\newtheorem{mytheorem}{Theorem}
\newtheorem{assumption}[mytheorem]{Assumption}
\newtheorem{lemma}[mytheorem]{Lemma}
\newtheorem{remk}[mytheorem]{Remark}

\newcommand{\hide}[1]{}

\newcommand{\name}{VoltanaLLM\xspace}

\renewcommand{\emph}[1]{\textit{#1}}

\newcommand{\predictor}{EcoPred\xspace}

\newcommand{\governor}{EcoFreq\xspace}

\newcommand{\router}{EcoRoute\xspace}

\DeclareMathOperator{\ttft}{TTFT}
\DeclareMathOperator{\itl}{ITL}

\DeclareMathOperator{\sorted}{sorted}

\DeclareMathOperator{\freq}{freq}
\DeclareMathOperator{\argmin}{argmin}

\def\BibTeX{{\rm B\kern-.05em{\sc i\kern-.025em b}\kern-.08em
    ÍT\kern-.1667em\lower.7ex\hbox{E}\kern-.125emX}}
\begin{document}

\title{\name: Energy-Efficient and SLO-Aware Disaggregated LLM Serving via Adaptive Frequency Control and State-Space Routing}

\author{
    \IEEEauthorblockN{
        Jiahuan Yu\textsuperscript{1}, 
        Aryan Taneja\textsuperscript{1,*}, 
        Junfeng Lin\textsuperscript{2,*},  
        Minjia Zhang\textsuperscript{1}
    }
    \vspace{0.1cm}
    \IEEEauthorblockA{\textsuperscript{1}\textit{Siebel School of Computing and Data Science, University of Illinois Urbana-Champaign}, Champaign, USA \\
    \{jiahuan2, aryant2, minjiaz\}@illinois.edu}
    \vspace{0.1cm}
    \IEEEauthorblockA{\textsuperscript{2}\textit{Department of Precision Instrument, Tsinghua University}, Beijing, China \\
    linjf21@mails.tsinghua.edu.cn}
    \thanks{\textsuperscript{*}Equal contribution.}
}


\maketitle


\begin{abstract}


The energy cost of Large Language Model (LLM) inference is rapidly becoming a barrier to sustainable and scalable deployment. Although modern serving architectures expose distinct prefill and decode behaviors, existing systems fail to exploit these phase differences for energy-efficient serving under strict latency SLOs. 
This paper introduces \name, the first system that explicitly targets and reduces the energy bloat in modern prefill-decode (P/D) disaggregated LLM serving. 
Guided by a control‑theory perspective, \name separates two levers: per‑instance operating‑point selection (GPU frequency per iteration) and system‑level state‑space routing of requests. We empirically observe that LLM inference exhibits a U‑shaped energy-frequency curve creating ``sweet spots'' that depend on phase behavior and load.
\name exploits this by combining phase‑specific, iteration‑level frequency selection driven by a lightweight, online‑adaptive latency predictor, with a decode state‑space guided router that avoids architectural granularity-induced inefficiencies, all while meeting desired SLOs.
We implement \name using SGLang and evaluate it across multiple models and real‑world workloads.
Our results show \name reduces end‑to‑end energy by up to 36.3\% versus a static max‑frequency baseline while maintaining high SLO attainment, and generalizes to newer GPUs.
These results point to sustainable LLM serving via phase‑aware, iteration‑level frequency selection coupled with architecture‑aware routing.
Source code is available in \texttt{\url{https://github.com/Supercomputing-System-AI-Lab/VoltanaLLM}}.

\end{abstract}

\section{Introduction} 
\label{sec:intro}

Large Language Models (LLMs) have become the cornerstone of modern AI services, powering applications ranging from conversational assistants~\cite{chatgpt,gemini}, code generation tools~\cite{copilot,cursor}, and agentic systems~\cite{zhang2025webpilot,wu2024autogen}.
Recent studies show that LLM inference now accounts for over 90\% of total AI infrastructure utilization for major providers~\cite{desislavov2023trends}.
As these systems are deployed at massive scale, inference energy consumption has emerged as one of the most critical bottlenecks to sustainable and cost-effective AI operation.
Recent industry reports indicate that even major cloud providers are constrained by power delivery limits, not just hardware availability~\cite{elsworth2025measuring,patel2024characterizing}.
This industry-wide inflection highlights a fundamental tension between performance scaling and energy capacity: data centers can no longer rely on simply provisioning more accelerators to sustain model growth, which motivates energy-efficient LLM serving as a primary design goal.

\begin{figure}[!t]
    \centering
    \includegraphics[width=\columnwidth]{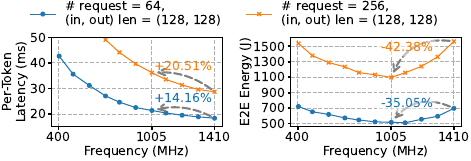}
    \caption{The decreasing per-token latency and the U-shaped energy-frequency curve across varying GPU frequencies, for LLaMA-3.1-8B~\cite{grattafiori2024llama} served on an A100 with SGLang~\cite{zheng2024sglang}.}
    \label{fig:u-shape-coupled}
\end{figure}

While the conventional wisdom of dynamic frequency scaling-based energy control holds that reducing frequency directly lowers power~\cite{gonzalez1997supply}, the net effect on energy (time $\times$ power) for LLM serving is not straightforward, as the increase in latency, which is crucial for user experience, varies significantly depending on workload characteristics.
In particular, our empirical profiling of LLM inference
reveals an interesting non-monotonic energy-frequency relationship.
As shown in \fref{fig:u-shape-coupled}, while reducing A100 GPU frequency from 1410 MHz to 1005 MHz ($-28.7\%$) does increase execution time, this increase is sub-linear.
Consequently, the total energy follows a U-shaped curve with respect to frequency.
This U-shaped trend suggests that at low frequencies, execution time dominates energy, whereas at high frequencies, power dominates; in the middle lies an energy sweet spot.

While this observation exposes a promising optimization opportunity, unfortunately what works well for conventional data center frequency control does not directly translate to energy-efficient LLM serving.
LLM applications are often real-time and interactive, with phase-specific \textit{Service Level Objectives} (SLOs), including \textit{Time-To-First-Token} (TTFT) and \textit{Inter-Token Latency} (ITL).
Violating these SLOs degrades user experience and downstream system responsiveness such as agentic pipelines~\cite{zhang2025webpilot}.
Modern LLM serving systems commonly employ continuous batching~\cite{yu2022orca} to batch prefill and decode tokens together for higher GPU utilization.
However, this coupled design introduces latency contention and phase interference~\cite{zhong2024distserve}, making it difficult to control energy without violating distinct, phase-specific SLOs.
Furthermore, achieving energy savings without SLO violations is complicated, and our analysis reveals several insights:
\emph{Insight-1}: Real-world workloads exhibit strong temporal variations in request rates and types, which lead to shifting prefill-decode throughput demand and short-term workload fluctuations.
For effectiveness, an energy controller must be adaptive and responsive (e.g., able to react on timescales of tens of milliseconds), but most static and power-capping methods incur coarse, high-latency adjustments (\sref{subsec:temporal-variation}).
\emph{Insight-2}: The energy-latency trade-off is neither uniform nor monolithic.
The prefill and decode phases have different U-shaped energy-frequency relationships and latency-frequency sensitivities.
This implies that a single, coupled frequency policy is inherently suboptimal and makes it difficult to meet distinct, phase‑specific SLOs (\sref{sec:obs-u-shape}).
\emph{Insight-3}: Energy efficiency also depends on architectural execution granularity.
We observe that when decode batch sizes cross certain thresholds (e.g., 256), GPU kernel tiling shift discontinuously, leaving compute units idle and creating a ``staircase'' pattern in latency and energy curves (\sref{subsec:tile-effect}).

To address these challenges, we build upon prefill-decode (P/D) disaggregation~\cite{patel2024splitwise,zhong2024distserve} and present \name, 
the first energy-efficient SLO-aware disaggregated LLM serving system that jointly optimizes GPU frequency and request routing under latency SLOs.
Specifically, \name introduces three key techniques:
(1) \governor (\sref{subsec:governor}), a responsive frequency controller that adjusts GPU frequency at fine-grained, per-iteration granularity within inference engine to respond to instantaneous workload fluctuations while adhering to SLOs;
(2) \predictor (\sref{subsec:predictor}), a lightweight yet accurate model inference time predictor, trained offline from profiling data and continuously adapted online to maintain accuracy under dynamic workloads and frequencies;
(3) \router (\sref{subsec:router}), an architecture-aware, state-space guided request router that dynamically steers decode requests by navigating a load-frequency state space to maximize overall energy efficiency while meeting SLOs.
Together, these techniques allow \name to jointly optimize for SLO attainment, energy efficiency, and dynamic load adaptation.

We implement \name on SGLang~\cite{zheng2024sglang}, a production-grade LLM inference engine, and make the following contributions:
\textbf{(1)} We propose the first optimization framework for energy efficient LLM inference under P/D disaggregation, which enables principled control of energy-latency trade-offs under SLO constraints.
\textbf{(2)} We are the first to identify two new architectural phenomena in disaggregated LLM serving: phase-specific U-shaped energy-frequency relationships and architectural granularity-induced inefficiency, which motivate phase-specific and architecture-aware energy control.
\textbf{(3)} We design and implement \name, the first SLO-aware and energy-efficient disaggregated LLM serving system that incorporates phase-specific and responsive frequency controller, as well as state-space guided request routing that delivers the best energy-latency efficiency.
We conduct extensive evaluations across various LLMs and workloads.
Our results show that \name achieves up to 36.3\% energy savings while preserving TTFT and ITL SLO attainment rates.


\section{Background and Related Work} 
\label{sec:bg}

\subsection{LLM Serving Systems} 
\label{subsec:bg-llm}



Numerous systems have been proposed to improve LLM serving efficiency, including advanced batching strategies for throughput optimization~\cite{fang2021turbotransformers,yu2022orca}, memory management techniques such as PagedAttention~\cite{kwon2023efficient}, CPU offloading~\cite{xu2024pie,sheng2023flexgen}, and vAttention~\cite{prabhu2025vattention}, and GPU kernel-level optimizations for accelerating attention and decoding~\cite{aminabadi2022deepspeed, dao2023flashattention2,fastertransformer,wang2021lightseq,shah2024flashattention}. To better utilize available resources and improve scheduling, model parallelism and pipelining frameworks have been introduced~\cite{aminabadi2022deepspeed,li2023alpaserve,pope2023efficiently}, along with parameter sharing mechanisms to reduce memory and computation overhead~\cite{sheng2023s}. Speculative decoding has emerged as a promising technique to reduce tail latency by predicting likely tokens ahead of time and verifying them in parallel~\cite{miao2024specinfer,stern2018blockwise}. Additionally, systems like FastServe~\cite{wu2023fast} explore preemptive scheduling policies to reduce job completion time (JCT) in multi-tenant serving environments. While these systems reduce energy consumption as a byproduct of latency or throughput improvements, we treat energy as a first-class optimization target and specifically exploit the control opportunities exposed by P/D-disaggregated serving. \name directly targets energy efficiency through frequency and routing-aware scheduling, which has been relatively underexplored despite its critical implications for sustainable deployment. 

\subsection{Energy-Efficient LLM Serving} 
\label{sec:bg-energy}

Several works have started to explore energy-efficient and power management frameworks for LLM inference~\cite{stojkovic2025dynamollm,stojkovic2025tapas,li2025ecoserve,reddy2025ai,fernandez2025energy,stojkovic2024towards}.
For example, DynamoLLM~\cite{stojkovic2025dynamollm} shows benefits of GPU frequency control in LLM inference serving based on request characteristics (predicted input/output tokens) to yield energy savings.
It proposes a hierarchical design to efficiently route incoming requests to respective GPU pools that vary by model sharding and frequency while changing frequency on a coarse-grained basis of load characteristics.
ThrottLL'eM~\cite{kakolyris2025throttll} predicts future KV cache usage and batch size to reduce frequencies and instance sizes for energy efficiency.
Similarly, $\mu$-Serve~\cite{qiu2024power} shows benefits of dynamic frequency scaling and proposes a model-serving framework to optimize for power consumption by co-serving multiple ML models.
In contrast, our approach is explicitly phase-aware: it distinguishes prefill and decode as separate control regimes with different latency sensitivities and energy behaviors, rather than applying a unified serving-time power policy.
Moreover, \name also performs per-engine-iteration frequency selection within the inference engine to react to short-term workload fluctuations under latency SLOs.

Taken together, these works show that DVFS and scheduling can improve energy and system efficiency, but they do not treat P/D disaggregation as an energy-control surface, nor do they study the phase-specific energy–frequency behavior and decode-side batch-boundary effects that motivate \name.
EcoServe~\cite{li2025ecoserve} focuses on reducing the operational and embodied carbon emissions throughout the hardware lifecycle. It proposes reusing, allocating and provisioning GPUs and CPUs based on offline/online inference, workload demand, model-size etc. TAPAS~\cite{stojkovic2025tapas} focused on routing requests to specific instances within rows/aisles of GPU data‑center racks by exploiting available thermal and power slack to avoid high-power hotspots. Recently, Heron\cite{reddy2025ai} proposes placing GPUs closer to renewable energy sources and adjust workloads for improved efficiency. These works are complementary to our contribution and underscore the importance of sustainable AI.
These systems operate primarily at the infrastructure or cluster level, whereas \name focuses on runtime operating-point selection.
We deeply study frequency-control under varying loads and \name is the first to explore energy-efficient LLM serving under P/D-disaggregated architecture systematically with SLO-aware feedback baked into the system.

\subsection{P/D Disaggregation in LLM Inference} 
\label{subsec:pd-dist}

To better manage the different compute characteristics of the prefill and decode phases, recent work has proposed prefill-decode (P/D) disaggregation, which assigns the two phases to separate GPU nodes. Systems such as SplitWise~\cite{patel2024splitwise}, TetriInfer~\cite{hu2024inference}, Llumnix ~\cite{sun2024llumnix} and DistServe~\cite{zhong2024distserve} show that this separation improves goodput and TTFT and ITL SLO attainment rates by avoiding phase-level contention and enabling distinct execution policies.
Our key distinction is that we treat P/D disaggregation not only as a latency-isolation mechanism, but also as an energy-control surface that exposes new opportunities for phase-specific adaptation.
As such, major open-source LLM libraries such as vLLM~\cite{kwon2023efficient} and SGLang~\cite{zheng2024sglang} have also added runtime support for this architectural pattern. However, these efforts focus primarily on improving metrics like latency and throughput.
In particular, prior P/D systems do not characterize the phase-specific U-shaped energy–frequency behavior or the decode-side batch-boundary inefficiencies that directly motivate our \governor and \router designs.
To our knowledge, \name is the first P/D-disaggregated LLM serving system, to jointly combine phase-specific per-engine-iteration frequency control, online-adaptive latency prediction, and architecture-aware routing for both energy and system efficiency.

\section{Observations and Opportunities} 
\label{sec:observations}

In this section, we characterize the performance-energy behavior of LLaMA-3.1-8B~\cite{grattafiori2024llama} served by SGLang~\cite{zheng2024sglang} on an A100 GPU.
We profile key metrics, including TTFT, ITL, and end-to-end (E2E) energy consumption (measured via NVIDIA Management Library~\cite{nvml}).
Furthermore, we analyze real-world workload traces, the Azure LLM Inference Trace 2024 dataset~\cite{stojkovic2025dynamollm}, to identify temporal and batching patterns.

\begin{figure}[t]
    \centering
    \begin{subfigure}[b]{0.49\columnwidth}
        \centering
        \includegraphics[width=\linewidth]{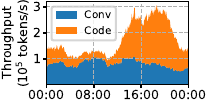}
        \caption{Prefill throughput.}
        \label{fig:pd-ratio-prefill}
    \end{subfigure}
    \hfill
    \begin{subfigure}[b]{0.49\columnwidth}
        \centering
        \includegraphics[width=\linewidth]{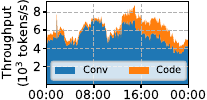}
        \caption{Decode throughput.}
        \label{fig:pd-ratio-decoding}
    \end{subfigure}
    \caption{Temporal variations of prefill and decode throughput in Azure LLM Inference Trace 2024 dataset. Conversation and code requests exhibit distinct diurnal patterns. Colors are stacked.}
    \label{fig:pd-ratio}
\end{figure}

\subsection{Multi-Timescale Workload Dynamics in LLM Serving}
\label{subsec:temporal-variation}

\begin{figure}[!t]
    \centering
    \begin{minipage}[b]{0.49\columnwidth}
        \centering
        \includegraphics[width=\columnwidth]{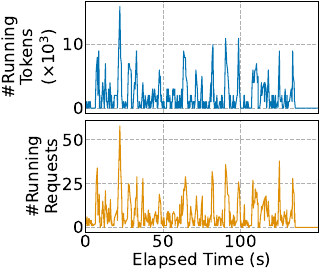}
        \caption{Rapid iteration-level workload fluctuations of running requests and  tokens of the prefill phase.}
        \label{fig:prefill-tokens-fluctuation}
    \end{minipage}
    \hfill
    \begin{minipage}[b]{0.49\columnwidth}
        \centering
        \includegraphics[width=\linewidth]{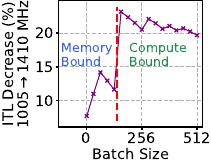}
        \caption{Decode ITL decrease percentage from frequency scaling (1005 $\to$ 1410 MHz) under various batch sizes.}
        \label{fig:decode-itl-diff}
    \end{minipage}
\end{figure}

Real-world LLM workloads exhibit variation at two distinct timescales.
At a \emph{coarse-grained} level, request arrival rates and prompt lengths fluctuate significantly over time, leading to shifting prefill-decode throughput demand and motivating energy adaptation.
Taking the Azure LLM Inference Trace 2024 dataset~\cite{stojkovic2025dynamollm} as an example, it categorizes requests into two types: \emph{conversation} and \emph{code}.
As shown in \fref{fig:pd-ratio-prefill}, the prefill throughput of conversation requests remains relatively stable over a day, while code requests exhibit a significant diurnal pattern, peaking in the afternoon and evening.
Furthermore, code requests typically have shorter decode lengths.
Consequently, as depicted in \fref{fig:pd-ratio-decoding}, the overall variation of the decode throughput is much smaller than that of prefill.
At a \emph{fine-grained} level, the number of active tokens and requests in prefill changes rapidly over time, as shown in \fref{fig:prefill-tokens-fluctuation}, due to dynamic batching, request completion, and various request lengths, which create short-term load fluctuations that directly affect GPU utilization and latency.
These P/D demand variations motivate phase-specific and workload-aware energy adaptation.
Most importantly, these fast dynamics also make static or coarse adjustments ineffective, highlighting the need for more responsive frequency control for energy efficiency.

\textbf{\underline{Insight \#1}}:
Real-world LLM inference workloads vary across two timescales: coarse-grained shifts in prefill-decode demand and fine-grained, iteration-level fluctuations in prefill batch composition.
These multi-timescale dynamics make static or coarse frequency control policies ineffective, motivating a phase-specific and responsive approach for energy-efficient LLM serving with SLO guarantees.


\subsection{U-Shaped Energy-Frequency Relationship and Sensitivities in P/D Disaggregated LLM Serving} \label{sec:obs-u-shape}


Frequency scaling is a common technique for energy optimization, yet its effect on LLM inference is non-monotonic.
Our measurements in \fref{fig:u-shape-coupled} reveal a clear U-shaped energy-frequency relationship, which indicates an energy-efficiency sweet spot.
To understand how this behavior manifests under P/D disaggregation, we  examine the two phases separately.

\begin{figure}[!t]
    \centering
    \begin{subfigure}[b]{0.49\columnwidth}
        \centering
        \includegraphics[width=\linewidth]{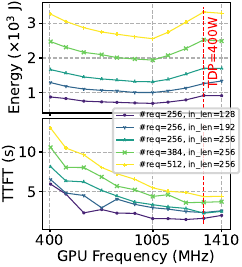}
    \end{subfigure}
    \hfill
    \begin{subfigure}[b]{0.49\columnwidth}
        \centering
        \includegraphics[width=\linewidth]{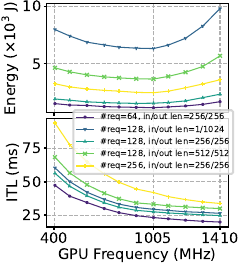}
    \end{subfigure}
    \begin{subfigure}[b]{0.49\columnwidth}
        \centering
        \includegraphics[width=\linewidth]{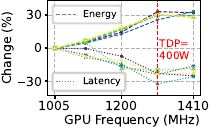}
        \caption{Prefill phase.}
        \label{fig:u-shape-p}
    \end{subfigure}
    \hfill
    \begin{subfigure}[b]{0.49\columnwidth}
        \centering
        \includegraphics[width=\linewidth]{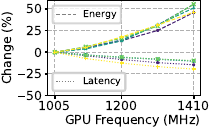}
        \caption{Decode phase.}
        \label{fig:u-shape-d}
    \end{subfigure}

    \caption{Impact of GPU frequency on energy (top) and latency (middle) for P/D phases on A100 with LLaMA-3.1-8B.
    The bottom shows the energy and latency relative changes by frequency increases from 1005 MHz to 1410 MHz.
    Prefill hits TDP limitations near 1305 MHz.
    Both phases exhibit U-shaped energy-frequency curves with 1005 MHz as energy-optimal points, but they differ significantly in shape and sensitivity.}
    
    \label{fig:u-shape}
\end{figure}

\fref{fig:u-shape} shows the energy and latency characteristics of the two phases across various GPU frequencies, using requests with varied input and output lengths for each phase.
Both phases show monotonically decreasing latency-frequency curves and U-shaped energy-frequency curves, with 1005 MHz consistently being the energy-optimal point before a sharp energy upsurge towards the maximum frequency (1410 MHz).
Frequencies below 1005 MHz are strictly suboptimal, as they increase both energy and latency.
However, the two phases differ markedly in shape and sensitivity, stemming from fundamental architectural factors.

According to~\cite{gonzalez1997supply}, the power $P$ of CMOS circuits (i.e., the physical foundation of GPUs) comprises a frequency-irrelevant static part and a frequency-dependent dynamic part:
\begin{equation}
    P = P_{\textbf{static}} + P_{\textbf{dynamic}} \approx P_{\textbf{static}} + CV^2f \sim A + B f^{1+\alpha},
\end{equation}
where $A$, $B$, $C$, $\alpha$ are coefficients, $V$ is voltage, which also increases with frequency.
We denote this change as $V^2\sim f^\alpha$, where $\alpha>0$ and can increase with $f$.
Thus, the execution time $T$ and energy consumption $E$ can be expressed as:
\begin{align}
    T &\sim f^{-\beta} \sim \left[(P-A)/B\right]^{-\beta/(1+\alpha)}, \label{eq:t-alpha-beta} \\
    E &= PT \sim A f^{-\beta} + B f^{1+\alpha - \beta}, \label{eq:u-shape} 
\end{align}
where $0\le\beta\le1$ is an arithmetic intensity-related coefficient: 
$\beta=1$ indicates fully compute-bound, and $\beta=0$ indicates fully memory-bound.
The $E$-$f$ plot of \eref{eq:u-shape} is a U-shaped curve, as the $A f^{-\beta}$ term dominates at low frequencies and the $B f^{1+\alpha - \beta}$ term dominates at high frequencies.
This theoretical analysis matches our empirical observation in \fref{fig:u-shape}.
Furthermore, for the compute-bound prefill phase, $\beta\approx1$, which means it benefits almost proportionally from higher frequency (\eref{eq:t-alpha-beta}) and gets significant TTFT reduction at the cost of higher energy (\fref{fig:u-shape-p} bottom), quickly driving the GPU toward its Thermal Design Power (TDP) limit at $\approx$1305 MHz.
In contrast, for the common memory-bound decode phase, $\beta<1$, which means decode derives a sub-linear benefit from higher frequency (\eref{eq:t-alpha-beta}), as \fref{fig:u-shape-d} bottom shows:  increasing $f$ from 1005 MHz to 1410 MHz yields only $\approx$20\% ITL reduction at the cost of $\approx$50\% higher energy.
Notably, as batch size grows and arithmetic intensity increases, decode gradually transitions toward compute-bound behavior and responds more strongly to higher frequencies (\fref{fig:decode-itl-diff}).

\textbf{\underline{Insight \#2}}: 
While both P/D phases exhibit U-shaped energy-frequency relationships and the corresponding energy sweet spots, they differ remarkably in shape and sensitivity.
Given that lowering frequency inevitably increases latency, these distinct characteristics motivate an adaptive, phase-aware policy to minimize energy while preserving SLOs.

\subsection{Architectural Granularity-Induced Energy Inefficiency}
\label{subsec:tile-effect}



\begin{figure}[t]
    \centering
    \begin{minipage}[b]{0.49\columnwidth}
        \centering
        \includegraphics[width=\linewidth]{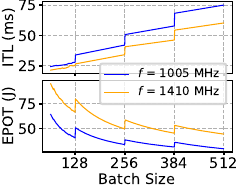}
        \caption{Batch size boundary and ``staircase'' effects in ITL and EPOT in the decode phase.}
        \label{fig:kernel-tile}
    \end{minipage}
    \hfill
    \begin{minipage}[b]{0.49\columnwidth}
        \centering
        \includegraphics[width=\columnwidth]{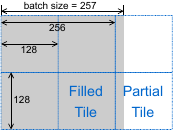}
        \caption{The tile quantization effect of GEMM. Tile size is 128, and request batch size is 257.}
        \label{fig:gemm-tile}
    \end{minipage}
\end{figure}

Modern LLM serving systems batch multiple requests to improve GPU utilization~\cite{yu2022orca}.
However, our profiling reveals that implicit batch size boundaries create architecturally induced inefficiencies in both energy and latency, especially during the decode phase.
As shown in \fref{fig:kernel-tile}, while increasing batch sizes generally reduces \textit{Energy-Per-Output-Token} (EPOT) and improves efficiency, this trend is periodically disrupted: crossing certain thresholds (e.g., 256 $\rightarrow$ 257) causes abrupt performance drops, resulting in a ``staircase'' pattern in ITL and EPOT.
A similar but weaker effect exists in prefill, which is masked by its larger token counts (see \appref{appendix:prefill-tile}).

This discontinuity arises from tile quantization.
GPUs execute GEMM by partitioning the output matrix into fixed-size ``tiles'' and assigning them to thread blocks~\cite{gemmtile}.
This ``staircase'' effect occurs when the batch size is not an exact multiple of the tile dimensions.
For example, as \fref{fig:gemm-tile} shows, when the batch size increases from 256 to 257, the GPU must launch an entirely new, but almost empty, tile (e.g., only 6.25\% useful data) to process this single extra request. 
This under-utilized tile requires the same execution time (duration) as a fully utilized tile. This leads to an increase in the total number of tiles, which in turn causes a sudden jump in latency (ITL) and a drop in efficiency (EPOT), forming the ``staircase’’ seen in \fref{fig:kernel-tile}.
While coupled serving architectures (e.g., chunked prefill~\cite{agrawal2024taming}) can mask this behavior by merging prefill and decode tokens into shared kernels, P/D disaggregated architecture amplifies the problem, as the decode instances operate with smaller, highly variable batch sizes. 

\textbf{\underline{Insight \#3}}: GPU architectural granularity, specifically SM-level tiling and scheduling boundaries, causes discontinuous ``staircase'' effects in energy and latency as batch sizes cross certain thresholds.
This hardware-level phenomenon is amplified in P/D disaggregated LLM inference, and reveals an overlooked source of energy inefficiency and motivates architecture-aware frequency and batching control.

\section{Problem Formulation} 
\label{sec:formulation}



We consider an LLM serving system that follows the P/D disaggregation architecture.
The system contains $N_P$ prefill instances (indexed by $p=1,\cdots,N_P$) and $N_D$ decode instances (indexed by $d=1,\cdots,N_D$), each running independently. 
Incoming requests must first complete prefill on one of the prefill instances and subsequently complete decode on one of the decode instances. Let $S_P$ and $S_D$ denote the TTFT and ITL SLO targets for these two phases.
The objective is to minimize the end-to-end energy consumption across all instances while meeting phase-specific SLOs: prefill $\ttft<S_P$, decode $\itl<S_D$.
This essentially yields the following constrained optimization problem: minimize total energy of all iterations across all instances, subject to satisfying TTFT and ITL SLO constraints for each request.

\section{System Design}
\label{sec:design}

Motivated by the insights in \sref{sec:observations}, we propose \name, an LLM serving system built on a P/D disaggregation architecture to achieve SLO-aware and energy-efficient inference.
The following subsections detail the design of \name.

\subsection{Design Principle: A Control Theory Perspective}
\label{subsec:control}




The global optimization problem in \sref{sec:formulation} defines the objective of minimizing end-to-end energy under per-phase latency SLO constraints. However, solving the problem exactly is combinatorial, and an NP-hard assignment problem~\cite{ross1975branch}, which is computationally intractable. Moreover, the future request arrivals are unknown at runtime. To make this problem tractable, we turn to a control theory perspective. Instead of solving the global problem monolithically, we observe that a P/D-disaggregated system naturally exposes two complementary control dimensions: (i) Instance-level execution configuration, which determines the latency-energy tradeoff of each prefill and decode instance. This is analogous to a local control loop in classical control system~\cite{wiener2019cybernetics}: each subsystem adjusts its internal operation point (e.g., frequency) based on its current measurable state. (ii) Assignment of incoming requests, which determines how load is distributed across instances. This corresponds directly to \emph{state-space control} in multi-subsystem systems~\cite{kalman1960general,kreindler1963contributions}: routing a request shifts an instance from one region of its state space to another, which affects both its future latency and its required operating point. 

These connections are not merely conceptual. In control theory terms, instance configuration control is a continuous, per-engine-iteration control problem over a single subsystem, while routing is a discrete state-space navigation problem over multiple subsystems whose state jointly determine the global energy-latency envelop of the system. This mapping guides our design: \name applies instance-level control to continuously select energy-efficient operating points, while global routing navigates the state spaces of instances to avoid high-energy regions such as architectural tile boundaries.

\subsection{Overall Architecture}
\label{subsec:design-overall}

\begin{figure}[t]
    \centering
    \includegraphics[width=\columnwidth]{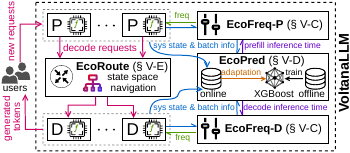}
    \caption{Overall architecture of \name. New requests are first handled by prefill instances (round-robin) to generate the first output token. \router then dispatches the requests to decode instance, from which generated tokens are streamed back to the users. \governor runs in a separate process to control the frequency of each P/D instance. \predictor provides prefill and decode model inference time predictions based on system states and batch information, while also collecting these metrics from all instances for online adaptation.}
    \label{fig:overall-arch}
\end{figure}

\fref{fig:overall-arch} illustrates the overall architecture of \name, which integrates three co-designed components that collectively deliver SLO-aware, energy-efficient LLM serving.

\noindent
\textbf{\governor} (\sref{subsec:governor}):
A phase-specific and iteration-level frequency controller for both prefill and decode instances based on current system state and batch information, maximizing energy efficiency while preserving SLO attainment. Its frequency selection is guided by the inference time predictor \predictor.

\noindent
\textbf{\predictor} (\sref{subsec:predictor}):
A accurate yet lightweight inference time predictor for both prefill and decode. Built on XGBoost~\cite{chen2016xgboost}, it combines both offline profiling and online adaptation to maintain accuracy under workload dynamics and mitigates distribution shift between offline and online data.

\noindent
\textbf{\router} (\sref{subsec:router}):
An energy-efficient router that assigns prefill-generated requests to decode instances.
It mitigates architectural granularity-induced energy inefficiency (\sref{subsec:tile-effect}) by analyzing and navigating each decode instance state spaces before making routing decisions, avoiding unnecessary transition into high-frequency regions.

\subsection{\governor: SLO-Aware and Iteration-Level Frequency Control for Responsive Energy Adaptation} 
\label{subsec:governor}


Although modern GPUs (e.g., NVIDIA A100/H100) expose APIs to set frequencies, they do not tune frequencies dynamically in response to load~\cite{vspetko2021dgx}.
As such, using them effectively for SLO-aware and energy-efficient LLM serving is non-trivial due to two challenges.
First, temporal variations in P/D demands (\sref{subsec:temporal-variation}) require distinct frequency policies for each phase, as they exhibit different optimal operating points depending on the workload.
The inevitable energy-latency trade-off (\sref{sec:obs-u-shape}) also requires a control policy that adapts to real-time workload variations while guaranteeing SLO attainment.
Second, coarse-grained frequency adjustment is ineffective in handling rapid iteration-level dynamism in workloads (\sref{subsec:temporal-variation}), yet existing fine-grained frequency control introduces expensive overhead.
Recent studies report $\sim$50 ms latency per change via blocking \texttt{nvidia-smi} commands\cite{stojkovic2025dynamollm}, which is unsuitable for latency-sensitive engine loops where each generation step typically completes within tens to hundreds of milliseconds.

\begin{figure}[!t]
    \centering
    \includegraphics[width=\columnwidth]{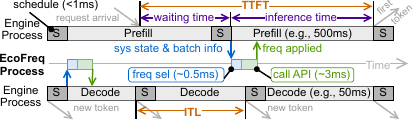}
    \caption{Execution flow of \governor with P/D instances. \governor runs as a separate lightweight process that collects system states and batch  from the engine and performs sub-millisecond iteration-level frequency control.}
    \label{fig:governor-exec-flow}
\end{figure}

To overcome these challenges, we design \governor, a responsive, SLO-aware controller that performs iteration-level GPU frequency adjustment for both prefill and decode instances. 
\fref{fig:governor-exec-flow} illustrates how \governor interacts with P/D instances.
To overcome the overhead of frequency settings via \texttt{nvidia-smi}, \governor introduces multiple optimizations.
It runs as a \textit{independent} process outside the critical inference path to completely hide the switching overhead, communicating with the inference engine via low-overhead ZeroMQ~\cite{zeromq}. 
Rather than expensively setting frequency via \texttt{nvidia-smi}, \governor leverages \texttt{nvmlDeviceSetGpuLockedClocks} API from NVIDIA Management Library~\cite{nvml} to apply new frequency at the whole-GPU granularity.
Our measurements show it consumes $<$3 ms.
Upon scheduling a new batch, the inference engine sends its current batch information $\mathbf{B}$ and system state $\mathbf{M}$ to \governor.
\governor then selects an appropriate GPU frequency and applies it immediately. 
This per-iteration loop enables responsive adaptation to instantaneous workload conditions.
Furthermore, the frequency selection algorithm (we will discuss it later) executes in $\sim$0.5 ms.
The total end-to-end overhead of \governor per iteration is thus $<$4 ms and is overlapped with the engine's model inference.
Consequently, although the beginning of an inference step may use the previous iteration's frequency, this period is negligible compared to the total model inference time (e.g., 500 ms for prefill and 50 ms for decode), introducing negligible control inaccuracy.
This lightweight design achieves per-iteration responsiveness to workload variations with negligible overhead, enabling \governor to apply per-engine-iteration, independent control for each phase and instance, 
thereby maximizing energy savings while ensuring SLO attainment.


The following steps details the SLO-aware, phase-specific frequency selection algorithm of \governor (\aref{alg:ecofreq}).
Let $\mathcal{F}$ be the set of frequency options.
Each invocation of \governor comprises three steps:
\ding{202} \textbf{Queue Check} (\alref{algline:step1-beg}-\ref{algline:step1-end}):
If there are requests in \emph{waiting} state (queued requests awaiting scheduling), 
\governor selects the highest frequency $\max(\mathcal{F})$ to promptly clear waiting requests and prevent SLO violations due to queue buildup. 
This prevents a scenario where a lower frequency, while sufficient for \emph{running} (actively executing decode steps) request SLOs, would reduce overall throughput, causing waiting requests to accumulate and subsequently violate their SLOs.
\ding{203} \textbf{Phase-Specific Adjustment} (\alref{algline:step2-beg}-\ref{algline:step2-end}):
The target token latency for each phase can be decomposed as:
\begin{equation}
    \text{Latency} = T_{\text{fixed}} + T_{\text{inference}}(f) \le \text{SLO}
\end{equation}
where $T_{\text{fixed}}$ is frequency-\emph{irrelevant} time (e.g., request scheduling and waiting) and $T_{\text{inference}}(f)$ is the frequency-\emph{dependent} GPU model inference time.
As shown in \fref{fig:governor-exec-flow}, $T_{\text{fixed}}$ differs by phase.
For \emph{decode}, $T_{\text{fixed}}$ is negligible scheduling time ($<$1ms).
Conversely, for \emph{prefill}, $T_{\text{fixed}}$ is primarily the request's waiting time (i.e., from arrival to being scheduled for running), which is non-negligible.
Since frequency scaling only controls $T_{\text{inference}}(f)$, \governor sets a phase-adjusted target $S$ for $T_{\text{inference}}(f)$ to ensure SLO attainment:
\begin{equation}
    T_{\text{inference}}(f) \le S = \begin{cases}
        S_P - \max\left( \mathbf{T}_{\text{waiting}} \right), & \text{prefill,} \\
        S_D, & \text{decode,}
    \end{cases}
\end{equation}
where $S_P$ and $S_D$ are TTFT and ITL SLOs, respectively, and $\max\left( \mathbf{T}_{\text{waiting}} \right)$ is the maximum waiting time among requests within a prefill batch.
\ding{204} \textbf{Frequency Selection} (\alref{algline:step3-beg}-\ref{algline:step3-end}):
Using the prefill ($T_P(\cdot)$) or decode ($T_D(\cdot)$) inference time predictors, \governor iterates through candidate frequencies $\mathcal{F}$ in ascending order, choosing the lowest frequency that makes $T_{\text{inference}}(f) \le S$.
If none meets the target, it defaults to $\max(\mathcal{F})$ to preserve SLO attainment.
The predictors $T_P(\cdot)$ and $T_D(\cdot)$ take $\sim$0.5 ms to execute each time.
We achieve this through careful model selection (discussed in \sref{subsec:predictor}), which enables per-iteration frequency selection of \governor.



\begin{algorithm}[!t]
\caption{\governor: SLO-Aware Frequency Selection}
 \label{alg:ecofreq}
 \begin{algorithmic}[1]

    \Require Current system status $\mathbf{M}$, batch information $\mathbf{B}$, frequency option list $\mathcal{F}$, model inference time predictors $T_P(\cdot)$ and $T_D(\cdot)$, TTFT SLO $S_P$, ITL SLO $S_D$.
    
    \Ensure Selected GPU frequency $f$.

    \If{$\Call{HasWaitingReqs}{\mathbf{M}}$}  \Comment{\textbf{step \ding{202} Queue check}} \label{algline:step1-beg}
        \State \Return $\max(\mathcal{F})$ \Comment{clear backlogged reqs timely}
    \EndIf \label{algline:step1-end}

    \If{$\Call{IsPrefill}{\mathbf{B}}$} \Comment{\textbf{step \ding{203} Phase adjustment}} \label{algline:step2-beg}
        \Statex /* prefill: deduct the waiting time from SLO budget */
        \State $S \gets S_P - \max( \mathbf{T}_{\text{waiting}}(\mathbf{B}) )$, $T_{\text{inference}}(\cdot) \gets T_P(\cdot)$ 
    \Else
        \State $S \gets S_D$, $T_{\text{inference}}(\cdot) \gets T_D(\cdot)$ \Comment{/* decode */}
    \EndIf \label{algline:step2-end}



    \For{$f \in \sorted(\mathcal{F})$}  \Comment{\textbf{step \ding{204} Frequency selection}} \label{algline:step3-beg}
        \If{$T_{\text{inference}}(\mathbf{M}, \mathbf{B}, f) \le S$}
            \State \Return $f$ \Comment{minimum frequency to meet the SLO}
        \EndIf
    \EndFor

    \State \Return $\max(\mathcal{F})$  \Comment{no freq can meet SLO, return max} \label{algline:step3-end}

 \end{algorithmic}
\end{algorithm}

The effectiveness of \governor relies on the accurate model inference time predictors $T_P(\cdot)$ and $T_D(\cdot)$ under varying system states and GPU frequencies.
To achieve this, we design online-adaptive predictors \predictor built on XGBoost~\cite{chen2016xgboost}, a model widely adopted for its ability to model performance while remaining lightweight and robust to feature interactions~\cite{chen2018tvm,chen2018learning}.
\predictor combines \emph{offline profiling} with \emph{online adaptation} to remain accurate under dynamic workloads.

\begin{figure}[t]
    \centering
    \begin{subfigure}[b]{0.49\columnwidth}
        \centering
        \includegraphics[width=\linewidth]{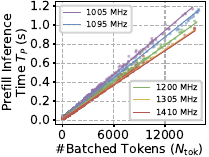}
        \caption{Prefill.}
        \label{fig:pred-ttft}
    \end{subfigure}
    \hfill
    \begin{subfigure}[b]{0.49\columnwidth}
        \centering

        \includegraphics[width=\linewidth]{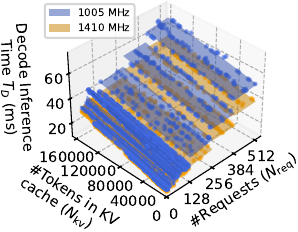}
        \caption{Decode.}
        \label{fig:pred-itl}
    \end{subfigure}
    \caption{Offline profiling results for LLaMA-3.1-8B on A100. Each point is a collected sample. Prefill latency grows near linearly with batched tokens, and decode latency exhibits tile-structured behavior across request numbers and KV-cache size. Lines and planes are visual guides to illustrate predictability. }
    \label{fig:pred-tile}
\end{figure}

\noindent
\textbf{Rationale: predictability and feature selection.}
For offline analysis, we perform the same profiling as in \sref{sec:obs-u-shape} but collect additional system state metrics: number of running requests (i.e., batch size) $N_{\text{req}}$, batched token number $N_{\text{tok}}$, and token number in KV cache storage $N_{\text{kv}}$.
\fref{fig:pred-tile} shows an example of profiling results for LLaMA-3.1-8B on A100.
The results clearly demonstrate that model inference time exhibits strong predictability under varying GPU frequencies and system states.
For prefill, inference time shows a strong linear relationship with $N_{\text{tok}}$. Conversely, decode inference time exhibits the ``staircase'' effect (as discussed in \sref{subsec:tile-effect}), yet remains highly predictable from $(N_{\text{req}}, N_{\text{kv}})$ within each tile.
These observations are formalized as:
\begin{align}
    T_P \left(\mathbf{M}, \mathbf{B}, f \right) \approx & a_f \cdot N_{\text{tok}} + b_f, \label{eq:pred-ttft} \\
    T_D \left( \mathbf{M}, \mathbf{B}, f  \right)  \approx & c_f \cdot N_{\text{req}} + d_f \cdot N_{\text{kv}} + e_f\ \ (\text{each tile}). \label{eq:pred-itl}
\end{align}
This predictability reflects the distinct computational characteristics of two phases.
For the compute-bound prefill phase, total computation theoretically scales linearly with the batched token number $N_{\text{tok}}$~\cite{vaswani2017attention,dao2023flashattention2}, yielding the highly linear relationship in \fref{fig:pred-ttft}.
In contrast, for the decode phase, $N_{\text{req}}$ determines the GEMM computational load~\cite{vaswani2017attention}, while $N_{\text{kv}}$ dictates the Attention layer computation and KV cache memory access~\cite{dao2023flashattention2}.
This profiling also captures the transition of decode from being memory-bound to compute-bound (as shown in \fref{fig:decode-itl-diff}).
As \fref{fig:pred-itl} illustrates, the latency gap between 1005 MHz and 1410 MHz is small for small batch sizes $N_{\text{req}}$, but widens significantly when batch sizes are larger as the workload becomes more computation-intensive.

Based on these observations, we build \predictor based on XGBoost~\cite{chen2016xgboost} to capture the predictability and create accurate models for $T_P(\cdot)$ and $T_D(\cdot)$:
\begin{align}
    T_P \left(\mathbf{M}, \mathbf{B}, f \right) &= \text{XGBoost}(f, N_{\text{tok}}),\\
    T_D \left(\mathbf{M}, \mathbf{B}, f \right) &= \text{XGBoost}(f, N_{\text{req}}, N_{\text{kv}}).
\end{align}
To ensure this offline model can be applied across diverse online serving situations, we profile uniformly over the feasible ranges of $N_{\text{tok}}$, $N_{\text{req}}$, and $N_{\text{kv}}$. This broad, distribution-agnostic sampling avoids bias toward any particular workload pattern and provides coverage of the model's execution, while online adaptation later corrects for any distribution shift.

\noindent
\textbf{Online adaptation to distribution shift.} While offline profiling employs uniform distributions to minimize bias, real deployments vary due to dynamic request arrivals and sequence lengths. \fref{fig:dist-shift} illustrates this clear distribution shift between offline and online data for a decode instance.

\subsection{\predictor: Online-Adaptive, Load-Aware Frequency–Latency Prediction for Iteration-Level Control} 
\label{subsec:predictor}

\begin{figure}[!t]
    \centering
    \begin{minipage}[b]{0.40\columnwidth}
        \centering
        \includegraphics[width=\textwidth]{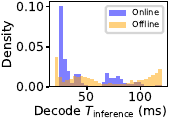}
        \caption{Distribution shift of decode model inference time between online and offline data.}
        \label{fig:dist-shift}
    \end{minipage}
    \hfill
    \begin{minipage}[b]{0.58\columnwidth}
        \centering
        \includegraphics[width=\textwidth]{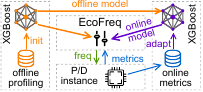}
        \caption{Overview of \predictor. The inference time prediction model is initialized by offline profiling and continuously refined using online metrics, enabling accurate runtime model inference time prediction.}
        \label{fig:predictor}
    \end{minipage}
\end{figure}

To overcome such distribution shift, we propose \emph{online adaptation}.
As shown in \fref{fig:predictor}, before the deployment, \predictor first performs the offline profiling and model training to build the initial offline predictors $T_P^{\text{offline}}(\cdot)$ and $T_D^{\text{offline}}(\cdot)$.
During runtime, \predictor continuously collects online performance metrics and initiates a background finetuning process every $X$ new samples to produce the updated finetuned models $T_P^{\text{online}}(\cdot)$ and $T_D^{\text{online}}(\cdot)$.
This online finetuning allows \predictor to maintain accuracy despite the dynamic workload variations.

At runtime, \governor queries \predictor to estimate P/D model inference times.
Since the models are lightweight, this prediction incurs negligible overhead (e.g., $<$0.5 ms). 

\subsection{\router: State Space-Guided Routing} 
\label{subsec:router}

While \governor adaptively controls GPU frequency for energy efficiency, request routing offers another key opportunity for energy optimization in multi-instance systems.

\begin{figure}[!t]
    \centering
    \includegraphics[width=\linewidth]{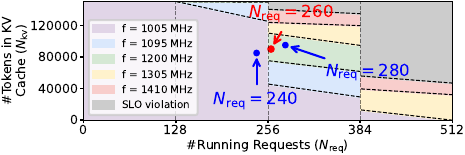}
    \caption{Example state space of a decode instance. The axes represent the number of running requests ($N_{\text{req}}$) and the number of tokens in KV cache ($N_{\text{kv}}$). Colored regions show the operating frequency $f$ chosen by \governor. The annotated points illustrate how different routing choices (e.g., $N_{\text{req}}=240, 260, 280$) place the instance in low- or high-frequency regions.}
    \label{fig:router-motivation}
\end{figure}

\noindent
\textbf{State-space perspective.}
As established earlier in \sref{subsec:predictor}, the ITL and energy consumption of a decode instance are functions of the triple $(N_{\text{req}}, N_{\text{kv}}, f)$.
Since \governor determines $f$ based on $(N_{\text{req}}, N_{\text{kv}})$, each instance's operating condition can be represented as a point in the $(N_{\text{req}}, N_{\text{kv}})$ state space.
When requests arrive, finish, or generate new tokens, the instance moves to a different point in this state space. 
A visualization of this state space (\fref{fig:router-motivation}) shows the ``frequency cliff'' created by batch-size boundaries (\sref{subsec:tile-effect}). When $N_{\text{req}}$ crosses a batch size boundary (e.g., 256), ITL increases significantly, which forces \governor to raise frequency sharply to maintain SLOs. This produces a disproportionate jump in energy consumption. Therefore, an effective routing policy should try to keep instances away from these high-cost regions whenever possible. 

\noindent
\textbf{Motivating example.}
Consider 2 decode instances, 520 total requests, and a batch boundary at 256.
A ``symmetric'' router (red point in \fref{fig:router-motivation}) dispatching 260 requests to each instance forces both to cross the ``cliff'', compelling higher frequency to meet the SLO.
In contrast, an ``asymmetric'' routing decision (blue points in \fref{fig:router-motivation}) could assign 240 requests to one instance, keeping it below the ``cliff'' at a low frequency, while the other instance handles 280 requests, incurring a small, SLO-compliant ITL increase. Although slightly imbalanced, this assignment is more energy efficient overall. This illustrates why routing should consider where each instance lies in the state space, rather than treating all instances uniformly. 

\begin{figure}[!t]
    \centering
    \includegraphics[width=\columnwidth]{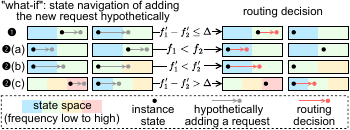}
    \caption{Illustration of \router for a system with 2 decode instances. Colors indicate the state space regions (mapped to different frequencies by \governor). The initial instance frequencies are $f_1$ and $f_2$, while $f^\prime_1$ and $f^\prime_2$ are the resulting frequencies after hypothetically adding the new request. \router evaluates these outcomes and selects the routing decision based on the corresponding decision rule (\ding{202}, \ding{203}a, \ding{203}b, \ding{203}c).}
    \label{fig:router-design}
\end{figure}

\noindent
\textbf{State-space guided routing design.}
We propose \router to exploit this opportunity.
Instead of exhaustively searching for the optimal assignment, \router makes online routing decision when each decode request arrives.
Leveraging the concept of state space, \router performs a ``what-if'' analysis: it hypothetically assigns the request to \emph{all} instances, navigates their state spaces and evaluates frequency changes, then makes the routing decision.
The algorithm compares the \emph{current} frequencies $\mathcal{F}$ with the hypothetically \emph{resulting} frequencies $\mathcal{F}^\prime$, leading to the following cases:
\ding{202} \textbf{Partial instances increase the frequencies, and the spread of $\mathcal{F}^\prime$ remains within the threshold} ($\max(\mathcal{F}^\prime)-\min(\mathcal{F}^\prime) \le \Delta$).
\router selects the instance with the lowest \emph{unchanged} frequency.
This rule avoids unnecessarily crossing the previously discussed frequency ``cliff''.
\ding{203} \textbf{Other situations}, covering (a) no changes ($\mathcal{F}=\mathcal{F}^\prime$), (b) all instance increase the frequencies, and (c) partial instance increase the frequency, but $\max(\mathcal{F}^\prime)-\min(\mathcal{F}^\prime) > \Delta$.
\router selects the instance with the lowest resulting frequency ($\min(\mathcal{F}^\prime)$) using round-robin.
\appref{subsubsec:router-algo-blk} gives a formal detailed version of above algorithm, and \fref{fig:router-design} provides an illustration for a 2-instance system.
It is noted that \router is also lightweight and does not introduce significant latency.
Its primary overhead comes from the latency prediction of \predictor, which takes $<$0.5 ms per query (\sref{subsec:predictor}).
Also, multiple queries for different decode instance candidates are batched together and processed by the underlying XGBoost model of \predictor to further minimize the latency.
Additionally, the routing algorithm itself relies on simple logic.
Consequently, \router introduces a negligible extra latency of $<$1 ms per request.






\begin{figure}[!t]
    \centering
    \includegraphics[width=0.49\columnwidth]{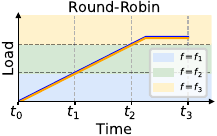}
    \includegraphics[width=0.49\columnwidth]{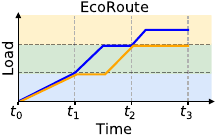}
    \caption{Conceptual comparison of round-robin and \router with two decoding instances.
    Line colors represent different instances. Shaded regions show the frequency levels required to meet SLOs ($f_1<f_2<f_3$). Under \router, one instance (orange) remains in a lower-frequency region for longer, yielding better energy efficiency than round-robin. }
    \label{fig:router-example}
\end{figure}

\fref{fig:router-example} conceptually illustrates how \router obtains energy benefits over a round-robin policy.
Round-robin suffers from the frequency ``cliff'' problem at batch size boundaries, like the ``symmetric'' router discussed before.
In contrast, \router is an ``asymmetric'' router, avoiding unnecessary frequency increases near the batch size boundaries.
Consequently, one instance (orange) operates for significantly more time at the energy-efficient low frequency than under the round-robin.


While \router is effective for decode instances, we apply a simple round-robin policy for prefill instances for two reasons.
First, the prefill phase does not exhibit the significant batch size boundary effect seen in decode, as its batched token numbers are typically large (\sref{subsec:tile-effect} and \appref{appendix:prefill-tile}).
Second, the prefill phase is largely stateless, showing little system status correlation between consecutive batches (discussed in \fref{fig:prefill-tokens-fluctuation}).
Consequently, the state-space navigation approach used by \router is inapplicable to prefill.

\section{Evaluation} 
\label{sec:eval}

\begin{figure*}[!t]
    \centering
    \includegraphics[width=\linewidth]{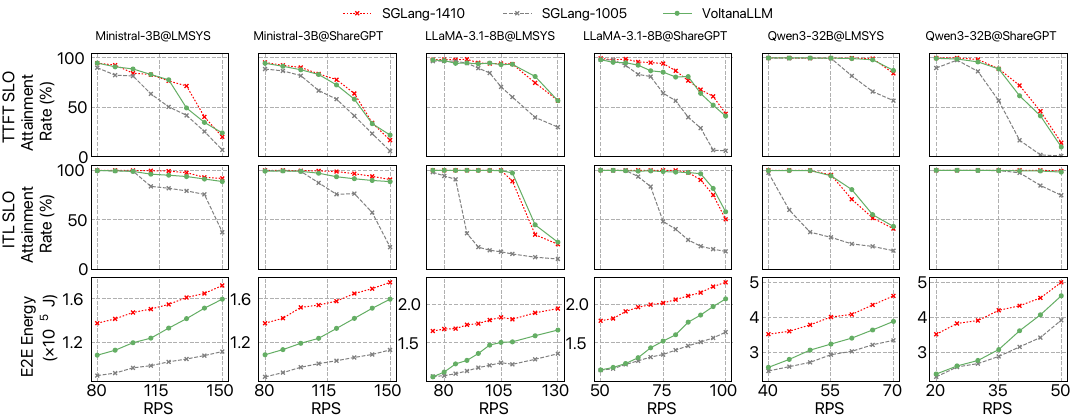}
    \caption{The TTFT/ITL SLO attainment rates and E2E energy consumption comparing \name and baselines across diverse models and datasets. \name consistently maintains latency SLO attainment rates comparable or slightly better than SGLang-1410, while significantly reducing E2E energy consumption.}
    \vspace{-1em}
    \label{fig:main-result}
\end{figure*}

\subsection{Evaluation Methodology}
\label{subsec:eval-method}

\noindent
\textbf{Implementation.}
We implement \name in Python on top of SGLang (v0.4.7.post1)~\cite{zheng2024sglang}, a widely used LLM inference framework with production-level P/D disaggregation support.
As described in \sref{subsec:governor}, \governor runs in a separate process to overlap overhead, and pyNVML~\cite{pynvml} is used for minimal-overhead frequency setting. 
\appref{subsubssec:xgboost} details the XGBoost model configurations of \predictor.

\noindent
\noindent
\textbf{Models and Workloads.}
We evaluate \name on Ministral-3B~\cite{ministral}, LLaMA-3.1-8B~\cite{grattafiori2024llama}, and Qwen3-32B~\cite{yang2025qwen3} using ShareGPT~\cite{sharegpt} and LMSYS-Chat-1M~\cite{zheng2023lmsys} workloads (more details in \appref{appendix:dataset-length}) with Poisson-distributed controlled request arrival rates (\emph{Requests-Per-Second}, RPS).

\noindent
\textbf{Metrics.}
We measure performance via TTFT, ITL, and the SLO attainment rate.
Following DistServe~\cite{zhong2024distserve}, SLO is defined as the target upper bounds of TTFT or ITL, and attainment rate refers to the percentage of requests satisfying these bounds ($\text{TTFT}\le S_P$, $\text{ITL}\le S_D$).
We report end-to-end energy consumption (Joules) measured via pyNVML.

\noindent
\textbf{Hardware Testbeds.} All experiments are performed on NVIDIA A100-80G SXM4 GPUs connected via NVLink.

\subsection{Main Result}
\label{subsec:main-result}

We evaluate \name across all models and datasets, using a 2P2D (2 prefill and 2 decode instances) configuration. 
For Qwen3-32B, 2-way Tensor Parallelism~\cite{shoeybi2019megatron} is used for each instance.
The frequency options of \governor are set to $\mathcal{F}=\{1005, 1410\}$ MHz, and the imbalance prevention threshold of \router is set to $\Delta=500$, which allows both frequencies to co-exist for different instances.
TTFT/ITL SLOs are set to 200/20, 600/60, 1200/120 ms for Ministral-3B, LLaMA-3.1-8B, and Qwen3-32B, respectively.
While DynamoLLM~\cite{stojkovic2025dynamollm} and throttLL’eM~\cite{kakolyris2025throttll} are the most relevant systems to our work, they are not included for comparison due to the lack of publicly available code.
As such, we construct two baselines based on SGLang:
\textbf{SGLang-1005}, using a static 1005 MHz frequency (the ``sweet spot'' described in \sref{sec:obs-u-shape}), and \textbf{SGLang-1410}, using the static maximum frequency (1410 MHz, the default strategy of A100~\cite{vspetko2021dgx}).
These baselines use SGLang's default round-robin request routing.
For each configuration, we report: (1) TTFT SLO attainment rate, (2) ITL SLO attainment rate, and (3) E2E energy consumption.

\fref{fig:main-result} shows our results, from which we have two key findings.
First, \name consistently achieves comparable TTFT SLO attainment rates and comparable or slightly better ITL SLO attainment rate compared to SGLang-1410, which operates at the static maximum frequency.
It significantly outperforms  SGLang-1005, which is energy efficient but yields substantially lower SLO attainment.
This ability to maintain high SLO attainment is attributed to the SLO-aware design of \governor.
It adaptively selects the GPU frequency for each engine iteration to ensure the model inference time remains within the budget.
Notably, the slightly better ITL SLO attainment compared to SGLang-1410 in some situations is attributed to \router.
As illustrated in \fref{fig:router-motivation}, \router allows one instance to avoid crossing a batch size boundary and frequency ``cliff'', at the cost of a marginal ITL increase in another instance.
This increase is outweighed by the ITL reduction benefit from avoiding the boundary.
Second, while preserving latency SLOs, \name reduces E2E energy consumption by up to 36.3\% compared to SGLang-1410.
The peak energy reduction occurs under the specific configuration of Qwen3-32B with ShareGPT at RPS of 20.
Its effectiveness is most pronounced at low request rates, where a static low frequency is sufficient for SLO targets, so \name dynamically operates at a low frequency most of the time.
Conversely, at high request rates, it dynamically increases frequency to maintain SLO attainment, resulting in smaller yet still significant energy benefits.
Detailed throughput indicate that \name achieves nearly the same throughput as SGLang-1410 at high request rates where more compute resources are required (more details are included \appref{subsubsec:ablation-throughput}).
Cumulative Distribution Functions (CDFs) of \name and baselines also show the same latency trends as SLO attainment rates in \fref{fig:main-result} (more details are included \appref{subsubssec:main-cdf}).

Overall, these improvements validate that SLO-aware frequency control and request routing effectively optimize energy consumption while preserving the quality of service.

\subsection{Analysis Results}
\label{subsec:analysis}

\begin{figure}[!t]
    \centering
    \includegraphics[width=\columnwidth]{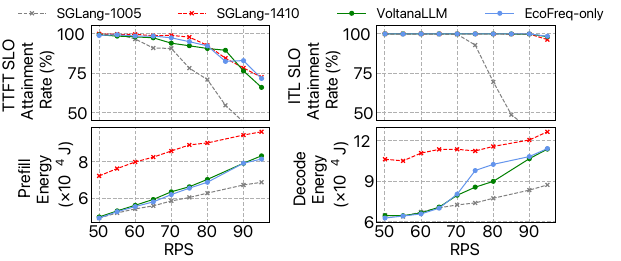}
    \caption{Comparison of \governor-only and full \name. \governor-only reduces energy for both prefill and decode while preserving SLOs, and 
    \router yields additional energy savings for the decode phase.} 
   
    \label{fig:sys-design-each-module}
\end{figure}

\begin{figure}[!t]
    \centering
    \includegraphics[width=0.95\columnwidth]{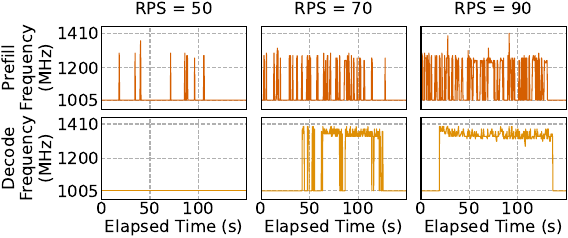}
    \caption{Real-time frequency traces of \governor-only for P/D instances at different request rates. At low RPS, instances primarily operate at low frequency, As RPS increases, both shift to high frequency more often.} 
    \label{fig:freq-time}
\end{figure}

\noindent
\textbf{How does each module of \name bring benefits?}
To quantify the per-component contribution in \name, we conduct experiments similar to \sref{subsec:main-result} using the ShareGPT dataset and LLaMA-3.1-8B model, but with two additional system configurations: \textbf{\governor-only} and a full version of \textbf{\name} (\governor+ \router).
\fref{fig:sys-design-each-module} presents the results.
First, \governor-only achieves substantial energy savings for both prefill and decode instances compared to the static high-frequency baseline SGLang-1410, validating the effectiveness of its SLO-aware frequency control policy.
\fref{fig:freq-time} provides an example of real-time frequency traces.
At low request rates, both P/D instances mainly operate at low frequency to conserve energy, while the proportion of high-frequency time increases for higher request rates due to SLO targets.
Second, the full \name achieves additional energy reductions compared to \governor-only, which is decode specific as \router is only applied to the decode instances.
\appref{subsubsec:router-example} gives an example to compares round-robin and \router by the frequencies of the two decode instances over time.
It shows \router can restrict the batch size of one instance under a specific boundary (256) so it can operate at a lower frequency.

\begin{figure}[!t]
    \centering
    \includegraphics[width=0.95\columnwidth]{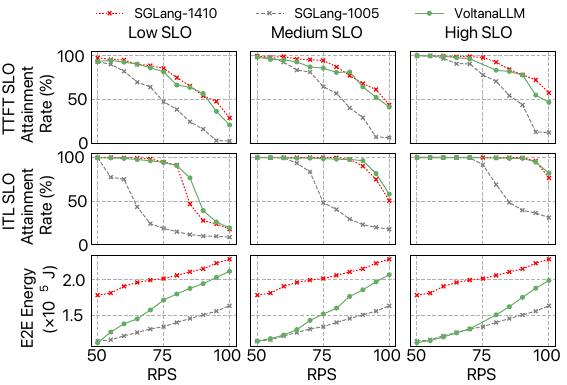}
    \caption{Comparison of \name and baselines under different SLO profiles (low, medium, high).
    Across all settings, \name maintains strong TTFT and ITL SLO attainment rates while providing significant energy savings.}
    \label{fig:slo-sensitivity}
\end{figure}

\noindent
\textbf{How does \name perform under different latency SLO targets?}
We evaluate \name on LLaMA-3.1-8B with the ShareGPT dataset, following the settings in \sref{subsec:main-result} but testing three different TTFT/ITL SLO profiles: 400/40 (``low''), 600/60 (``medium''), and 800/80 (``high'').
\fref{fig:slo-sensitivity} presents the results.
Across all SLO profiles, \name consistently achieves SLO attainment rates comparable or slightly better than the static maximum frequency baseline.
Specifically, as the SLO constraints are relaxed (i.e., from ``low'' to ``high''), \name's adaptive frequency control trades higher latency for lower energy consumption.
This demonstrates the latency-energy trade-off, providing the flexibility to define custom scenario and phase-specific SLOs, such as prioritizing low ITL for responsiveness or low energy consumption to reduce cost.
This result further emphasizes the necessity of the phase-specific design as discussed in \sref{sec:observations}.


\begin{figure}[!t]
    \centering
    \includegraphics[width=0.95\columnwidth]{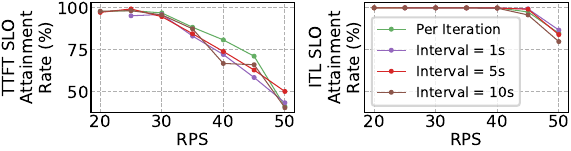}
    \caption{Latency SLO attainment rates across different frequency control intervals. Our per-iteration design provides the best responsiveness, and thereby the highest SLO attainment rates.}
    \label{fig:time-window}
\end{figure}

\noindent
\textbf{How important is it to have per-iteration frequency control?}
In \sref{subsec:temporal-variation}, we show rapid iteration-level workload variation in LLM serving, which necessitates the design for per-iteration frequency control.
\name provides this iteration-level responsiveness, unlike prior approaches~\cite{stojkovic2025dynamollm} that adjust frequency at coarser, window-based intervals (e.g., 5s).
To verify its benefit, we adapt \governor to operate at various fixed time intervals, evaluating it using the ShareGPT dataset on LLaMA-3.1-8B under the 1P1D configuration (where \router has no effect).
The results in \fref{fig:time-window} demonstrate that window-based frequency control degrades SLO attainment for both phases.
Specifically, per-iteration frequency responsiveness is particularly critical for the prefill phase, where window-based control shows larger degradation.
Due to dynamic batching, the total number of tokens processed in a prefill batch (and thus the optimal frequency) can vary significantly from one iteration to the next, as \fref{fig:prefill-tokens-fluctuation} shows.
A window-based approach is too coarse-grained to react to such rapid changes.
In contrast, as shown in \fref{fig:bs-time}, the load of the decode instance shows smoother change in batched requests, leading to relative insensitivity to window interval settings.

\begin{figure}[!t]
    \centering
    \includegraphics[width=0.85\columnwidth]{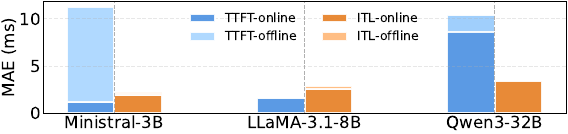}
    \caption{Mean absolute errors (MAEs) of \predictor for both TTFT and ITL, comparing the offline-only model to the online-adapted model.} 
    \label{fig:pred-accuracy}
\end{figure}

\noindent
\textbf{Is \predictor accurate? Can online adaptation improve accuracy?}
To verify these aspects, we compare \predictor's predicted values against the empirical latency results from \sref{subsec:main-result}, evaluating both the initial offline-trained and the online-adapted models.
\fref{fig:pred-accuracy} shows the results.
Both TTFT and ITL predictions consistently achieve low mean absolute errors (MAEs) across all models, which enables the SLO-aware frequency selection of \governor.
Furthermore, online adaptation provides a clear MAE improvement, which successfully mitigates the sample distribution shift (\fref{fig:dist-shift}).


\begin{figure}[!t]
    \centering
    \includegraphics[width=\columnwidth]{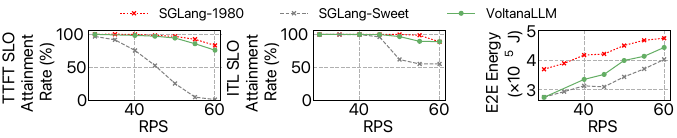}
    \caption{Latency SLO attainment and E2E energy consumption of \name on GH200, using Qwen3-32B and ShareGPT dataset.}
    \label{fig:exp-gh200}
\end{figure}

\noindent
\textbf{Can \name generalize to other hardware?}
To verify \name's generalizability, we evaluate it on NVIDIA GH200 Grace Hopper Superchips with Qwen3-32B (without Tensor Parallelism) and ShareGPT dataset.
GH200 exhibits similar but different U-shaped energy-frequency curves compared to A100: the energy sweet spot is 1095 MHz for prefill and 1395 MHz for decode, with a maximum frequency of 1980 MHz (more details in \appref{subsubssec:u-shape-gh200}).
Accordingly, we set \governor's frequency options as $\mathcal{F}_P=\{1095, 1980\}$ MHz and $\mathcal{F}_D=\{1395, 1980\}$ MHz.
The baselines are \textbf{SGLang-1980} (fixed maximum frequency) and \textbf{SGLang-Sweet} (P/D instances fixed at their respective sweet spots).
Other settings follow \sref{subsec:main-result}.
\fref{fig:exp-gh200} shows the results, which leads to a similar conclusion as on A100: \name maintains comparable latency SLO attainment and substantial energy savings, demonstrating its generalizability to different hardware.


\subsection{Comparison with Alternative Solutions and Parameters}
\label{subsec:further-analysis}

We further evaluate \name against several alternative configurations to assess its robustness and efficiency.
First, we compare \name with SGLang running at various intermediate static frequencies (1095, 1200, and 1305~MHz) and find that these static frequencies consistently consume more energy at low request rates or fail to meet SLOs at high request rates due to their lack of dynamic adaptability (detailed results in \appref{subsubsec:intermidate-static-freq}).
Similarly, a comparison against power-capped SGLang (e.g., 350W) reveals that static power limits strictly prevent the system from boosting frequency when necessary to maintain SLOs, while still failing to optimally lower frequencies during periods of low demand (see \appref{subsubsec:power-capped} for more details).
A detailed breakdown of phase-specific energy savings shows that while the decode phase offers greater gains at low request rates due to lower arithmetic intensity, the prefill phase contributes more significantly at high request rates by exploiting short-term, iteration-level low-load durations (more details in \appref{subsubsec:phase-specific-energy-gain}).
Evaluations with multi-level frequency options (e.g., 5 levels) demonstrate that while finer granularity can offer marginal energy savings for decode, it may slightly degrade SLO attainment rates (see \appref{subsubsec:exp-multi-level} for more details).
Finally, a sensitivity analysis of the imbalance threshold $\Delta$ in \router indicates that the system is generally robust to this parameter, though larger values can cause a slight negative impact on ITL SLO attainment under high loads (further analysis in \appref{subsubsec:delta-analysis}).

\section{Conclusions}
\label{sec:conclusion}



In this work, we present \name, the first system for energy-efficient and SLO-aware LLM inference under prefill-decode (P/D) disaggregation.
\name exploits the fundamentally different compute and memory behaviors of the two phases to introduce phase-specific, iteration-level frequency control and architecture-aware, state-space navigation-based routing.
Together, these techniques enable per-engine-iteration control of latency-energy that existing serving systems cannot achieve.
Through comprehensive evaluation on state-of-the-art LLMs and real-world datasets, \name achieves up to 36.3\% energy savings while preserving latency SLO attainment.
These results highlight the potential of fine-grained, phase-aware co-design in advancing sustainable LLM serving systems.

\textbf{Future research} will focus on extending this framework to non-GPU hardware accelerators, which requires hardware-specific calibration of energy-frequency curves and retraining the \predictor models.
Additionally, integrating \router with advanced learning-based routing policies presents a promising direction, especially considering the architectural granularity-induced inefficiencies identified in this study.

\section{Acknowledgments}

We gratefully acknowledge Alaa Youssef, Vijay Naik, and Yu Chin Fabian Lim from IBM for their valuable discussions on energy-efficient LLM inference. 
We thank Kartik Ramesh for assistance with conducting preliminary experiments and providing feedback on the manuscript during the early stage of this project, and Mingtao Hu for configuring the test environment on GH200.
This research was supported by the National Science Foundation (NSF) under Grant No. 2441601.
The work utilized the Delta and DeltaAI system at the National Center for Supercomputing Applications (NCSA) and Jetstream2 at Indiana University through allocation CIS240055 from the Advanced Cyberinfrastructure Coordination Ecosystem: Services \& Support (ACCESS) program, which is supported by National Science Foundation grants \#2138259, \#2138286, \#2138307, \#2137603, and \#2138296.
The Delta advanced computing resource is a collaborative effort between the University of Illinois Urbana-Champaign and NCSA, supported by the NSF (award OAC 2005572) and the State of Illinois.
UIUC SSAIL Lab is supported by research funding and gift from Google, IBM, Amazon, and AMD, including the Google ML and Systems Junior Faculty Award.

\clearpage

\bibliographystyle{IEEEtranS}
\bibliography{_s99_refs}

\clearpage
\appendix





\subsection{Batch Size Boundaries of Prefill Phase}
\label{appendix:prefill-tile}

\begin{figure}[!ht]
    \centering

    \includegraphics[width=\linewidth]{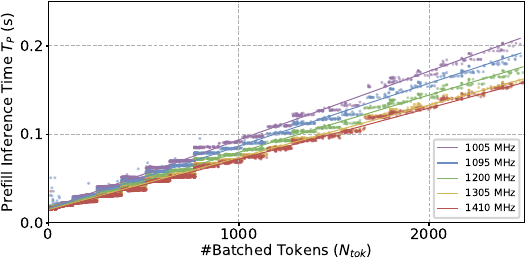}
    \caption{Zoomed-in version of \fref{fig:pred-ttft} for small batched token numbers: relationship between prefill model inference time and batched token numbers ($N_{\text{tok}}$) (LLaMA-3.1-8B on A100).}
    \label{fig:ttft_vs_tokens-small}
\end{figure}

\fref{fig:ttft_vs_tokens-small}, a zoomed-in view of \fref{fig:pred-ttft}, focuses on the prefill phase with small numbers of batched tokens to show the relationship between prefill model inference time and batched token numbers ($N_{\text{tok}}$).
We can observe similar ``staircase-like'' tile effect like decode phase (\fref{fig:pred-itl}).
However, this effect only exist when batched token numbers are relative small.
When number of batched tokens goes over about 2000 (which is the dominate situation), this effect gradually becomes less significant.

\subsection{Formal Routing Algorithm Description of \router}
\label{subsubsec:router-algo-blk}

\aref{alg:ecoroute} provides a detailed formal version of routing algorithm proposed in \sref{subsec:router}.
Check \sref{subsec:router} for text description of each step.

\begin{algorithm}[!ht]
 \caption{\router: Decode Request Routing}
 \label{alg:ecoroute}
 \begin{algorithmic}[1]

    \Require Decode instance list $\mathcal{D}=\{d_i\}_{i=1}^n$ and their state list $\mathcal{M}=\{m_i\}_{i=1}^n$, request for routing $r$, frequency decision of \governor $\freq(\cdot)$, frequency threshold $\Delta$.

    \Ensure Selected decode instance $d\in\mathcal{D}$.

    \State $\mathcal{F} \gets \{ \freq(m_i)\}_{i=1}^n$ \Comment{current freq of all instances}
    \State $\mathcal{F}^\prime \gets \{ \freq(m_i \oplus r)\}_{i=1}^n$ \Comment{frequency after hypothetically adding request $r$, $\oplus$ denotes adding request}

    \If{\Call{PartialChanged}{$\mathcal{F}$, $\mathcal{F}^\prime$} $\And$ $\max(\mathcal{F}^\prime)-\min(\mathcal{F}^\prime) \le \Delta$}
        \Statex /* \textbf{case} \ding{202}: if some but not all instances change the frequency, and the frequency difference $\le$ threshold */
        \State \Return $d_{\argmin_i \Call{Unchanged}{\mathcal{F}, \mathcal{F}^\prime}}$ \Comment{min unchanged}
    \EndIf
    \Statex /* \textbf{case} \ding{203}: other situations */
    \State \Return \Call{RoundRobin}{$d_{\argmin_i \mathcal{F}^\prime}$} \Comment{round-robin of min}


 \end{algorithmic}
\end{algorithm}

\subsection{XGBoost Configuration of \predictor}
\label{subsubssec:xgboost}

As discussed in \sref{subsec:predictor}, we employ XGBoost~\cite{chen2016xgboost} to predict prefill and decode model inference times based on system states and batch information.
The model configuration details are as follows:
\begin{lstlisting}
import xgboost as xgb

# prefill model
model_prefill = xgb.XGBRegressor(
    objective="reg:absoluteerror",
    n_estimators=100000,
    learning_rate=0.5,
    booster='gblinear',
    max_depth=6,
    subsample=0.8,
    colsample_bytree=0.8,
    tree_method='hist',
    random_state=42,
    eval_metric='mae',
    early_stopping_rounds=1000,
)

# decode model
model_decode = xgb.XGBRegressor(
    objective="reg:absoluteerror",
    n_estimators=1000,
    learning_rate=0.1,
    booster='gbtree',
    max_depth=6,
    subsample=0.8,
    colsample_bytree=0.8,
    tree_method='hist',
    random_state=42,
    eval_metric='mae',
    early_stopping_rounds=200,
)
\end{lstlisting}

\subsection{Length Distribution of ShareGPT and LMSYS Dataset}
\label{appendix:dataset-length}

\begin{figure}[!ht]
\includegraphics[width=\linewidth]{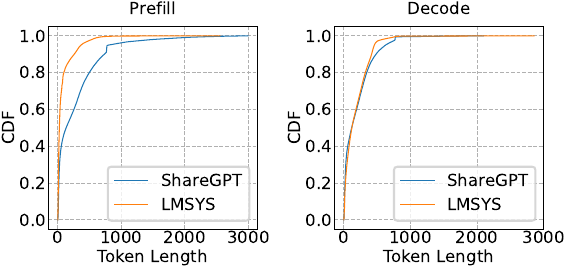}
    \caption{CDF of prefill and decode lengths for the ShareGPT and LMSYS datasets.}
    \label{fig:dataset-distribution}
\end{figure}

\tref{tab:length_statistics} summarizes the statistical properties of the ShareGPT and LMSYS datasets used in the evaluation, including the mean and standard deviation of their input and output lengths.
The cumulative distribution functions (CDFs) for these length distributions are visualized in \fref{fig:dataset-distribution}.
Analysis of this data shows that:

\begin{itemize}
    \item ShareGPT has longer prefill and decode length compared to LMSYS dataset.
    \item ShareGPT has larger prefill/decode length ratio.
\end{itemize}

\begin{table}[!ht]
    \centering
    \caption{Length distribution of datasets used in our evaluation. Std refers to standard deviation.}
    \label{tab:length_statistics}
    \begin{tabularx}{\columnwidth}{C | C C C C}
        \toprule
        \textbf{Dataset} & \multicolumn{2}{c}{\textbf{Prefill}} & \multicolumn{2}{c}{\textbf{Decode}} \\
        \cmidrule(lr){2-3} \cmidrule(lr){4-5}
        & \textbf{Mean (ms)} & \textbf{Std (ms)} & \textbf{Mean (ms)} & \textbf{Std (ms)} \\
        \midrule
        ShareGPT & 280.27 & 375.58 & 190.90 & 209.15 \\
        LMSYS    & 78.40  & 133.29 & 174.57 & 166.13 \\
        \bottomrule
    \end{tabularx}

\end{table}

\subsection{Extra Throughput Data of The Main Result}
\label{subsubsec:ablation-throughput}

\begin{figure}[!ht]
    \centering
    \includegraphics[width=\linewidth]{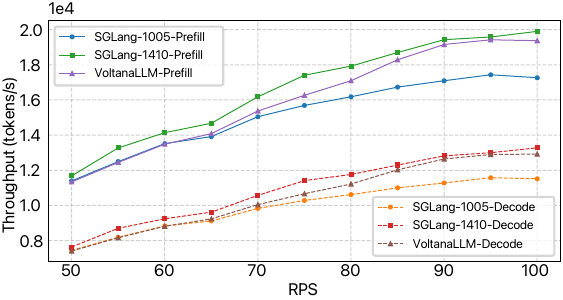}
    \caption{Comparing throughput of \name with baselines for LLaMA-3.1-8B and ShareGPT dataset.}
    \label{fig:throughput}
\end{figure}

We compute the throughput from the experimental results in \sref{subsec:main-result} using ShareGPT dataset on LLaMA-3.1-8B.
As shown in \fref{fig:throughput}, \name achieves slightly lower throughput than SGLang with static maximum frequency due to its opportunistic frequency scaling, but not at the cost of SLO violations.
Furthermore, at high request rates where greater throughput is required, \name can also adaptively achieve nearly the same throughput as SGLang with static maximum frequency.

\subsection{Extra TTFT and ITL CDFs of The Main Result}
\label{subsubssec:main-cdf}

\begin{figure*}[!ht]
    \centering
    \includegraphics[width=0.95\linewidth]{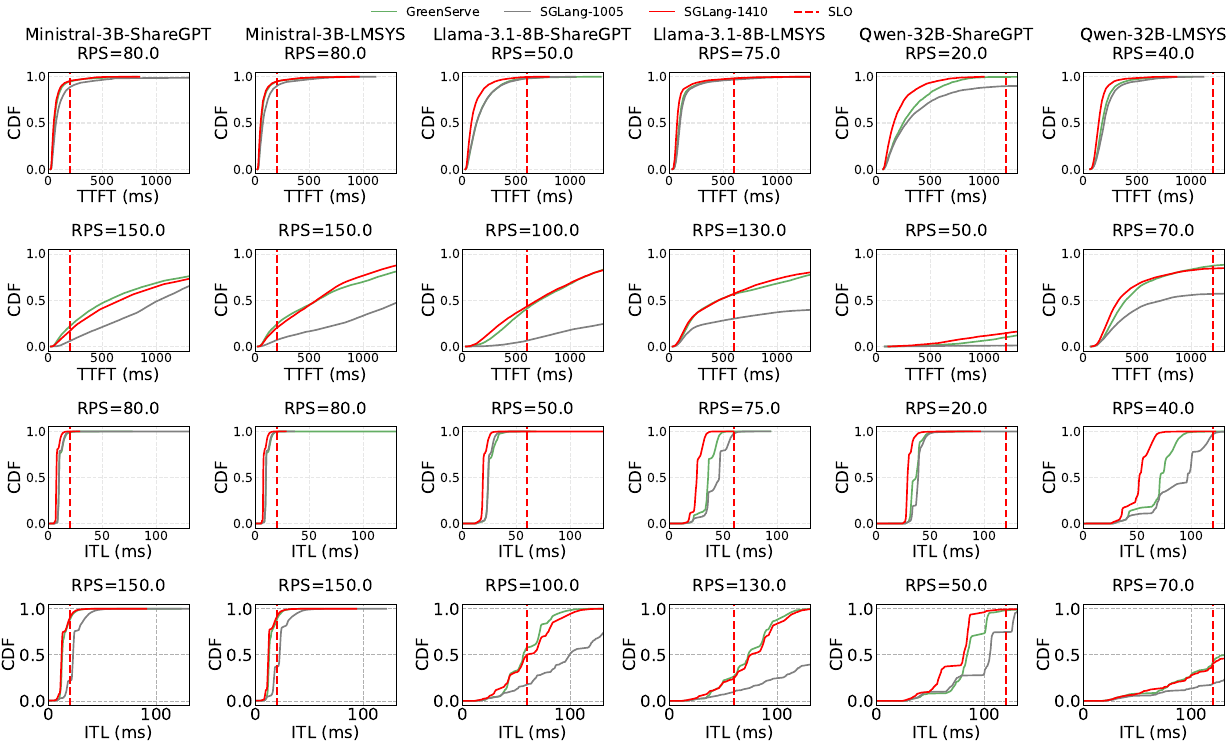}
    \caption{CDF of TTFT (top 2 rows) and ITL (bottom 2 rows) across various models and datasets (each column). For each model and dataset combination, we pick the smallest (row 1 and 3) and the largest (row 2 and 4) request arrival rates to shows the distinct behavior of \name under different workloads. Data comes from \fref{fig:main-result}.}
    \label{fig:main-cdf}
\end{figure*}

\fref{fig:main-cdf} shows the TTFT and ITL CDFs corresponding to the results in \sref{subsec:main-result}.
The figure provides CDFs across all model and dataset combinations (organized by column).
For each combination, CDFs are presented for both low request arrival rates (rows 1 and 3) and high request rates (rows 2 and 4).
The results show that when the request arrival rate is low, \name achieves latency CDFs similar to SGLang-1005, as the low frequency is sufficient for SLO attainment.
In contrast, when the request rate is high, \name adaptively increases its frequency, achieving latency CDFs similar to SGLang-1410.

\subsection{Comparison with More Intermediate Static Frequency Baselines}
\label{subsubsec:intermidate-static-freq}

\begin{figure}[!ht]
    \centering
    \includegraphics[width=\columnwidth]{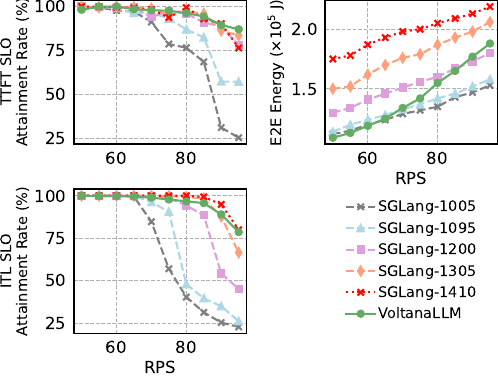}
    \caption{Comparison of energy consumption and SLO attainment between \name and SGLang with intermediate static frequencies across varying request rates, using LLaMA-3.1-8B model and ShareGPT dataset.}
    \label{fig:intermidate-static-freq}
\end{figure}

To evaluate whether intermediate static frequencies can achieve similar efficiency benefits, we compare \name against SGLang operating at fixed intermediate frequencies of 1095~MHz, 1200~MHz, and 1305~MHz, using LLaMA-3.1-8B model and ShareGPT dataset.
As \fref{fig:intermidate-static-freq} shows, because these static configurations inherently lack dynamic adaptability, they consistently consume significantly more energy than \name at low request rates.
Conversely, at high RPS, these static intermediate frequencies fail to provide the necessary compute throughput, which degrades SLO attainment compared to \name.
This confirms that static frequency tuning, regardless of the specific frequency chosen, is insufficient for energy-efficient, SLO-aware frequency control.

\subsection{Comparison with Power-Capped Baselines}
\label{subsubsec:power-capped}

\begin{figure}[!ht]
    \centering
    \includegraphics[width=\columnwidth]{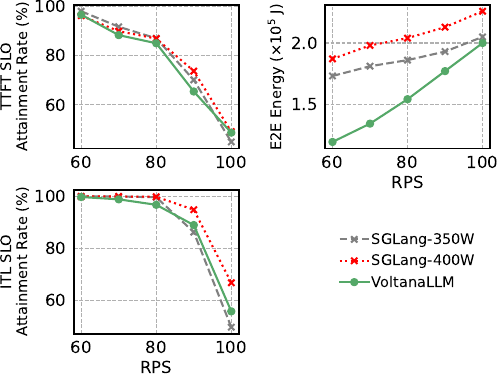}
    \caption{Comparison of energy consumption and SLO attainment between \name and a 350W power-capped SGLang across varying request rates, using LLaMA-3.1-8B model and ShareGPT dataset.}
    \label{fig:power-capped}
\end{figure}

We also evaluate \name against power-capped baselines.
Fundamentally, power-capping acts as an indirect upper-bound on frequency, rather than a responsive alternative to our per-engine-iteration frequency control.
We evaluate a 350W statically-capped SGLang configuration against the default 400W limitation, using LLaMA-3.1-8B model and ShareGPT dataset.
Our results in \fref{fig:power-capped} demonstrate that power-capping lacks adaptability for workload dynamism.
Specifically, at low request rates, the power-capped baseline fails to optimally lower frequencies, thus consuming substantially more energy than \name for similar SLO attainment.
Furthermore, at high RPS, the power cap strictly prevents the system from boosting frequency when necessary, which inherently degrades overall SLO attainment.

\subsection{Detailed Phase-Specific Energy Gain}
\label{subsubsec:phase-specific-energy-gain}

\begin{figure}[!ht]
    \centering
    \includegraphics[width=\columnwidth]{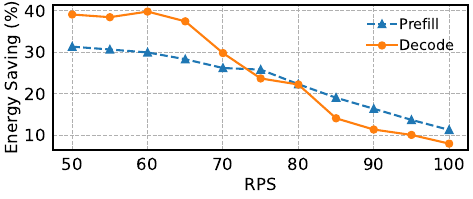}
    \caption{Energy saving percentage of each phase for \name under the LLaMA-3.1-8B@ShareGPT configuration in \sref{subsec:eval-method}.}
    \label{fig:phase-specific-energy-gain}
\end{figure}

To understand where our solution brings the most gain within the disaggregated architecture, we decompose the energy savings from the configuration LLaMA-3.1-8B@ShareGPT in \sref{subsec:eval-method} into the prefill and decode phases.
As \fref{fig:phase-specific-energy-gain} shows, the benefits are load-dependent.
At low request rates, the decode phase yields greater energy savings, which is attributed to its inherently lower arithmetic intensity.
Conversely, at high RPS, the prefill phase contributes more significantly to the overall energy savings. This shift occurs because our \governor effectively exploits short-term, low-load durations within the prefill phase to perform iteration-level frequency reductions, whereas the decode instances must remain at higher frequencies to satisfy strict ITL SLOs under heavy load.

\subsection{Do More Frequency Levels Bring Benefits?}
\label{subsubsec:exp-multi-level}

\begin{figure}[!ht]
    \centering
    \includegraphics[width=\columnwidth]{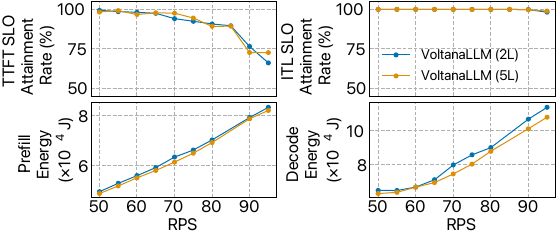}
    \caption{Comparison of \name with different frequency levels. \name (2L) uses two frequency levels, while \name (5L) uses five. Both configurations maintain similar TTFT/ITL attainment and energy trends when combined with \router.}

    \label{fig:freq-level}
\end{figure}

In \sref{subsec:main-result}, \name is configured with two-level frequency options.
To analyze the potential benefits of finer-grained levels, we extend \governor's frequency options to five levels: $\mathcal{F}=\{1005, 1095, 1200, 1305, 1410\}$ MHz, and evaluate it using the ShareGPT dataset on LLaMA-3.1-8B, following the same settings as \sref{subsec:main-result}.
\fref{fig:freq-level} presents the results.
For prefill, finer frequency granularity yields negligible benefits, as the significant iteration-level workload variation (\fref{fig:prefill-tokens-fluctuation}) makes the middle levels largely unused.
In contrast, the decode reveals a trade-off: finer granularity offers slight energy savings but at the cost of lower SLO attainment rates.
This allows users to select appropriate granularities based on their specific performance-energy priorities.

\subsection{Sensitivity Analysis of $\Delta$ in \router}
\label{subsubsec:delta-analysis}

As discussed in \sref{subsec:router}, $\Delta$ serves as a guardrail against extreme workload imbalance, where a smaller value indicates a stricter tolerance.
While $\Delta$ remains largely inactive for 2-level frequency configurations where the frequency gap is a deliberate design target, it becomes crucial in multi-level frequency settings (e.g., 5-level configurations).
In these scenarios, $\Delta$ constrains the frequency variance between instances, preventing \governor from over-allocating requests to high-frequency instances while leaving lower-frequency instances underutilized.

\begin{figure}[!ht]
    \centering
    \includegraphics[width=\columnwidth]{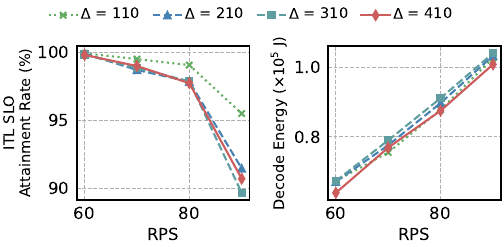}
    \caption{Impact of $\Delta \in \{110, 210, 310, 410\}$ on ITL SLO attainment and energy performance under 5-level frequency setting.}
    \label{fig:delta-analysis}
\end{figure}

To evaluate the sensitivity and robustness of \router to the frequency spread threshold $\Delta$, we set \name across $\Delta \in \{110, 210, 310, 410\}$, with 5-level frequencies $\mathcal{F}=\{1005, 1095, 1200, 1305, 1410\}$ MHz in \governor.
Other settings follow \appref{subsubsec:exp-multi-level}.
Our results shown in \fref{fig:delta-analysis} indicate that while larger $\Delta$ values can slightly degrade ITL SLO attainment at high RPS due to imbalance, the choice of $\Delta$ is generally robust, with all tested values delivering stable and comparable performance.

\subsection{An Example of Comparison between Round-Robin and \router}
\label{subsubsec:router-example}

\begin{figure}[!ht]
    \centering
    \includegraphics[width=\columnwidth]{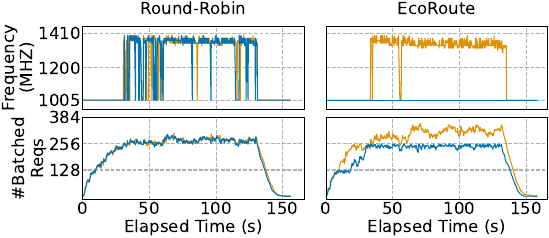}
    \caption{An example of how \router differs from round-robin policy (RPS=75). Each line represents a decode instance. \router keeps one instance below the batch size boundary (256), allowing it to run at a lower frequency for longer time. }    
    \label{fig:bs-time}
\end{figure}

\fref{fig:bs-time} compares round-robin and \router by the batch size and frequencies of the two decode instances over time.
\fref{fig:router-example} in \sref{subsec:router} is a conceptual version of it.
It shows \router can restrict the batch size of one instance under a specific boundary (256) so it can operate at a lower frequency.
That is a good example to show how \router save energy by avoiding frequency ``cliff'' discussed in \fref{fig:router-motivation}.

\subsection{U-Shape Figures of GH200}
\label{subsubssec:u-shape-gh200}

\begin{figure}[!ht]
    \centering
    \begin{subfigure}[b]{0.49\columnwidth}
        \centering
        \includegraphics[width=\linewidth]{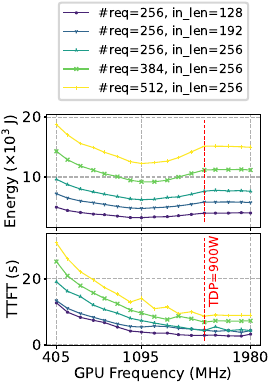}
        \caption{Prefill phase.}
    \end{subfigure}
    \hfill
    \begin{subfigure}[b]{0.49\columnwidth}
        \centering
        \includegraphics[width=\linewidth]{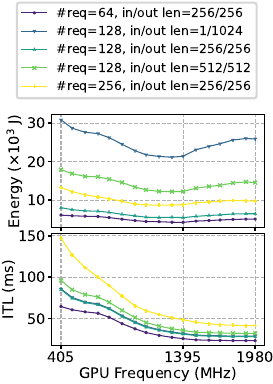}
        \caption{Decode phase.}
    \end{subfigure}
    \caption{Impact of GPU frequency on energy (up) and latency (down) for P/D phases on GH200 with Qwen3-32B. Prefill hits TDP limitations (900 W) near 1600 MHz.
    Both phases exhibit U-shaped energy-frequency curves, but prefill reach energy sweet spot $\approx$1095 MHz, while decode reach energy sweet spot $\approx$1395 MHz.}
    \label{fig:u-shape-gh200}

\end{figure}

\fref{fig:u-shape-gh200} illustrates the U-shaped energy-frequency curves and monotonically decreasing latency-frequency curves when serving Qwen3-32B~\cite{yang2025qwen3} on a GH200.
These results exhibit trends similar to those of the A100 (\fref{fig:u-shape}), but with distinct hardware-specific differences.
The prefill phase remains power-intensive, hitting the GH200's 900W TDP limitation at $\approx$1600 MHz. Furthermore, the two phases possess different energy sweet spots: the optimal point for prefill is $\approx$1095 MHz, while for decode, it is $\approx$1395 MHz.

\subsection{What happens in varying the P/D throughput demand ratio?}
\label{subsubsec:ablation-pd-ratio}

\begin{figure}[!ht]
    \centering
    \begin{subfigure}[t]{0.42\columnwidth}
        \centering
        \includegraphics[width=\linewidth]{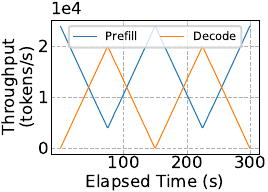}
    \end{subfigure}
    \hfill
    \begin{subfigure}[t]{0.53\columnwidth}
        \centering
        \includegraphics[width=\linewidth]{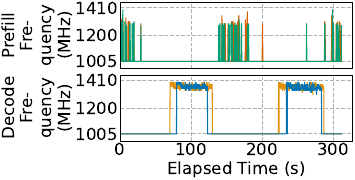}
    \end{subfigure}
    \caption{Left: P/D throughput with our synthetic dataset. Right: frequency dynamics of all instances in \name.}
    \label{fig:ablation-pd-ratio}
\end{figure}

As discussed in \sref{subsec:temporal-variation}, the temporal variation of P/D throughput demand ratio requires phase-specific frequency control.
However, evaluating the whole Azure LLM Inference Trace dataset incurs significant GPU cost and time given it only shows fluctuation on the scale of hours.
Hence, we use a synthetic dataset with P/D demand ratio fluctuating in 5 minutes.
\fref{fig:ablation-pd-ratio} shows the evaluation results with LLaMA-3.1-8B using the same configuration in \sref{subsec:main-result}.

The results shown in \tref{tab:ablation-pd-ratio} confirm the adaptiveness of \name.
It can still maintain comparable SLO attainment rates to SGLang at maximum frequency, while obtain significant energy savings.
\fref{fig:ablation-pd-ratio} shows the frequency dynamics.
When prefill throughput demand is high, the prefill instance operates at a higher frequency, while the decode instance remains at a low frequency.
Conversely, when decode throughput demand is high, the decode instance operates at a higher frequency, while the prefill instance remains at a low frequency.

\begin{table}[!ht]
    \centering
    \caption{Latency SLO attainment and end-to-end (E2E) energy consumption under a synthetic workload with varying P/D demand ratios.}
    \label{tab:ablation-pd-ratio}
    \begin{tabularx}{\columnwidth}{C | C C C}
        \toprule
        \textbf{Metrics} & TTFT SLO Attainment Rate (\%) & ITL SLO Attainment Rate (\%) &  E2E Energy (J) \\
        \midrule
        \textbf{\name} &96.05 &  91.74 & 289409\\
        \textbf{SGLang-1005} &77.86 & 36.78 & 257768 \\
        \textbf{SGLang-1410} &97.85 & 88.92 & 405340\\
        \bottomrule
    \end{tabularx}
\end{table}

\end{document}